\documentclass[prd,superscriptaddress,footinbib,twocolumn,floatfix]{revtex4}
\usepackage{amsmath,amssymb,amsfonts,bm,amscd}
\usepackage{graphics}

\makeatletter
\renewcommand{\@makecaption}[2]{
  \vskip\abovecaptionskip
  \sbox\@tempboxa{\small\sf #1: #2}%
  \ifdim \wd\@tempboxa >\hsize
  \small\sf #1: #2\par
  \else
    \global \@minipagefalse
    \hb@xt@\hsize{\hfil\box\@tempboxa\hfil}%
  \fi
  \vskip\belowcaptionskip}
\makeatother

\def\ba{\begin{eqnarray}}
\def\ea{\end{eqnarray}}

\def\frak{\mathfrak}

\def\tilde{\widetilde}

\def\bar{\overline}

\newcommand{\ft}[2]{{\textstyle\frac{#1}{#2}}}

\def\Dslash{\,\,{\raise.15ex\hbox{/}\mkern-12mu D}}
\def\Dbarslash{\,\,{\raise.15ex\hbox{/}\mkern-12mu {\bar D}}}
\def\delslash{\,\,{\raise.15ex\hbox{/}\mkern-9mu \partial}}
\def\delbarslash{\,\,{\raise.15ex\hbox{/}\mkern-9mu {\bar\partial}}}
\def\pslash{\,\,{\raise.15ex\hbox{/}\mkern-9mu p}}
\def\calDslash{\,\,{\raise.15ex\hbox{/}\mkern-12mu {\cal D}}}

\newcommand{\Z}{{\mathbb Z}}

\newcommand{\Q}{{\mathbb Q}}

\def\CB{{\mathcal B}}
\def\CC{{\mathcal C}}

\def\CF{{\mathcal F}}

\def\CH{{\mathcal H}}
\def\CI{{\mathcal I}}

\def\CM{{\mathcal M}}
\def\CN{{\mathcal N}}

\def\CT{{\mathcal T}}

\def\ft{{\mathfrak{t}}}

\def\fs{{\mathfrak{s}}}

\renewcommand{\bar}{\overline}

\begin{document}
\preprint{CALT-TH-2018-033}

\title{3d TQFTs from Argyres-Douglas theories}

\author{Mykola Dedushenko}
\affiliation{Walter Burke Institute for Theoretical Physics, California Institute of Technology, Pasadena, CA 91125, USA}

\author{Sergei Gukov}
\affiliation{Walter Burke Institute for Theoretical Physics, California Institute of Technology, Pasadena, CA 91125, USA}

\author{Hiraku Nakajima}
\affiliation{Kavli Institute for the Physics and Mathematics of the Universe (WPI), The University of Tokyo,
Kashiwa, Chiba 277-8583, Japan}

\author{Du Pei}
\affiliation{Walter Burke Institute for Theoretical Physics, California Institute of Technology, Pasadena, CA 91125, USA}
\affiliation{Center of Quantum Geometry of Moduli Spaces, Department of Mathematics,
 Aarhus University, DK-8000, Denmark}

\author{Ke Ye}
\affiliation{Walter Burke Institute for Theoretical Physics, California Institute of Technology, Pasadena, CA 91125, USA}

\begin{abstract}
We construct a new class of three-dimensional topological quantum field theories (3d TQFTs) by considering generalized Argyres-Douglas theories on $S^1 \times M_3$ with a non-trivial holonomy of a discrete global symmetry along the $S^1$. For the minimal choice of the holonomy, the resulting 3d TQFTs are non-unitary and semisimple, thus distinguishing themselves from theories of Chern-Simons and Rozansky-Witten types respectively. Changing the holonomy performs a Galois transformation on the TQFT, which can sometimes give rise to more familiar unitary theories such as the $(G_2)_1$ and $(F_4)_1$ Chern-Simons theories.  
Our construction is based on an intriguing relation between topologically twisted partition functions, 
wild Hitchin characters, and chiral algebras which, when combined together, relate
Coulomb branch and Higgs branch data of the same 4d $\CN=2$ theory. We test our proposal by applying localization techniques to the conjectural $\CN=1$ UV Lagrangian descriptions of the $(A_1,A_2)$, $(A_1,A_3)$ and $(A_1,D_3)$ theories.
\end{abstract}


\maketitle

\newcommand{\be}{\begin{equation}}
\newcommand{\ee}{\end{equation}}

\subsection*{Introduction}

In this letter we propose a new link between two subjects,
both of which have a certain degree of mystery associated to them.
One subject is the study of 3- and 4-manifolds via topological twists of 4d $\CN=2$ theories~\cite{Witten:1988ze}.
The other subject involves Argyres-Douglas theories~\cite{Argyres:1995jj,Argyres:1995xn}
whose spectrum of superconformal operators and their correlation functions
remain as a challenge even 20 years after the discovery of such interacting 4d SCFTs.

The interplay between these two subjects leads to yet another mystery observed in~\cite{Fredrickson:2017yka} --- a non-trivial relation between spectra of Higgs and Coulomb branch operators
in the same Argyres-Douglas theory. One of the goals of the present letter is to generalize this observation, which we believe will be a useful step toward a deeper conceptual understanding of spectra of SCFTs and their relation to quantum invariants of 3- and 4-manifolds.

In practice, for a (generalized) Argyres-Douglas theory $\CT$, we construct a 3d TQFT by choosing the four-dimensional spacetime to be
\be
M_4 \; = \; S^1 \times M_3.
\label{M3M4}
\ee
A general feature of an Argyres-Douglas theory is that it possesses a $\Z_N$ global symmetry, which can be used to turn on a non-trivial holonomy $\gamma \in \Z_N$ along the $S^1$ in the above geometry. Further, the theory is topologically twisted along $M_3$ by identifying the Riemannian holonomy group $\mathrm{Spin}(3)$ with $SU(2)_R\subset SU(2)_R \times U(1)_r$ of the R-symmetry of the 4d $\CN=2$ Argyres-Douglas theory, and by a standard argument the partition function is expected to be a topological invariant of $M_3$.

When the holonomy is co-prime to $N$, {\it i.e.}~$\gamma\in\Z_N^\times$, the 3d TQFT is expected to be semisimple, and as in the famous Turaev's construction \cite{Turaev},
this 3d TQFT is associated to a modular tensor category (MTC) determined by $\CT$ and $\gamma$,
\be
(\CT,\gamma) \quad \leadsto\quad  \CC_{\CT}^{\gamma}.
\ee
This MTC, in turn, can be determined either from the geometry of the wild Hitchin moduli space \cite{Fredrickson:2017yka}
or from the chiral algebra of \cite{Beem:2013sza}. Furthermore, the multiplicative action of $\Z_N^\times$ on $\gamma$ gives the action of the Galois group $\mathrm{Gal}\left(\Q(e^{2\pi i/N})/\Q\right)\simeq \Z_N^\times$ on $\CC_{\CT}^{\gamma}$.

Our proposal in this paper is closely related to several recent developments
which also express topologically twisted partition functions on 3-manifolds in terms of the SCFT data.
For example, it was conjectured in \cite{Dedushenko:2017tdw} that the Coulomb branch index of
a 4d $\CN=2$ theory, Lagrangian or not, is equal to the topological partition function on $S^1 \times S^3$,
\be
Z (S^1 \times S^3) \; = \; \CI_{\text{Coulomb}} ({\frak t}),
\ee
where ${\frak t}$ is the holonomy for a $U(1)$ subgroup of R-symmetry along the $S^1$. This is the diagonal subgroup, which we will denote as $U(1)_{\ft}$, of $U(1)_r\times U(1)_R$ with the second factor being the Cartan of $SU(2)_R$. 
Similarly, in~\cite{Gukov:2016gkn} it was noted that partially twisted partition functions of 3d $\CN=2$ theories
are conveniently encoded in the data of an MTC associated to that theory. For example, a twisted partition function on $S^1 \times \Sigma_g$ has the form
\be
Z_{\text{3d}} (S^1 \times \Sigma_g) \; = \; \sum_{\lambda} (S_{0 \lambda})^{2-2g},
\label{ZSSigma}
\ee
where the sum is over isomorphism classes of simple objects in the MTC, $0$ stands for the unit object, and $S_{0\lambda}$ are entries of the modular $S$ matrix which are proportional to quantum dimensions of the objects labeled by $\lambda$. 
When twisted partition functions are computed via localization --- which, of course, is only possible in Lagrangian theories ---
the sum on the right-hand side of~\eqref{ZSSigma} can be interpreted as a sum over Bethe vacua \cite{Nekrasov:2014xaa}, and
\be
S \; = \; \text{``handle gluing operator.''}
\label{Shandle}
\ee
The $T$ matrix in the MTC can also be computed via localization, by considering a non-trivial fibration of $S^1$ over $\Sigma_g$,
\be
T \; = \; \text{``fibering operator.''}
\label{Tfibering}
\ee
This interpretation was used in \cite{Dedushenko:2017tdw,Gukov:2016gkn} to compute simple topological partition functions
in Lagrangian and non-Lagrangian theories (see also \cite{Gukov:2015sna,Benini:2015noa,Benini:2016hjo,Gukov:2017zao,Closset:2017bse,Closset:2018ghr}).

In our present context, the MTC associated to a 4d $\CN=2$ Argyres-Douglas theory also
comes equipped with $S$ and $T$ matrices. Using these $S$ and $T$ matrices, we can conveniently write the topological partition function on a 4-manifold of the form \eqref{M3M4}, where $M_3$ is defined by an arbitrary plumbing graph, as
\be
Z_{\text{AD}} (S^1 \times M_3) = \sum_{\lambda_v} \prod_{\text{vertices}} S_{0 \lambda_v}^{2-\deg(v)} T_{\lambda_v \lambda_v}^{a_v} \prod_{\text{edges}} S_{\lambda_v \lambda_v'}.
\label{ZSTgeneral}
\ee
For example, the modular tensor category $\CC_{(A_1,A_2)}^{1}$ associated to the simplest Argyres-Douglas theory $(A_1,A_2)$ with the minimal $\Z_5$ holonomy is that of the $(2,5)$ Virasoro minimal model (a.k.a. Lee-Yang model), with $S$ and $T$ matrices given by
\be\label{ST}
S
=  \frac{2}{\sqrt{5}}
\begin{pmatrix}
- \sin \frac{2\pi}{5} & \sin \frac{\pi}{5} \\
\sin \frac{\pi}{5} & \sin \frac{2\pi}{5}
\end{pmatrix}
,\quad
T = 
\begin{pmatrix}
e^{\frac{11 \pi i}{30}} & 0 \\
0 & e^{- \frac{\pi i}{30}}
\end{pmatrix}.
\ee
The Lee-Yang model has $c = - \frac{22}{5}$ and two primaries of scaling dimensions $h = 0, - \frac{1}{5}$. Although it is one of the simplest non-unitary MTCs, it still contains interesting structures, such as a non-abelian anyon and a non-trivial associator. With $\gamma = 4$, the category $\CC_{(A_1,A_2)}^{4}$ is expected to be the complex conjugate of the Lee-Yang MTC, while $\CC_{(A_1,A_2)}^{2}$ and $\CC_{(A_1,A_2)}^{3}$ are respectively the unitary Fibonacci MTC and its conjugate, which can be realized by $(G_2)_1$ and $(F_4)_1$ Chern-Simons theories. They all have the same fusion rules, and the $F$ and $R$ matrices can be found, {\it e.g.}, in \cite[Sec.~5.3]{RSW}. 

\subsection*{MTCs from wild Hitchin moduli spaces}

Generalized Argyres-Douglas theories can be constructed by compactifying 6d (2,0) theory on $\mathbb{CP}^1$ with an irregular singularity --- or ``wild ramification'' --- and possibly another regular singularity \cite{Eguchi:1996vu,Bonelli:2011aa,Xie:2012hs,Wang:2015mra}. The moduli space of Coulomb branch vacua of such a theory on $S^1 \times \mathbb{R}^3$ is identified with the moduli spaces $\CM_H$ of wild Higgs bundles on $\mathbb{CP}^1$.

$\CM_H$ has many interesting properties (see {\it e.g.}~\cite[Sec.~2]{Fredrickson:2017yka} for a review from the viewpoint of Argyres-Douglas theories). Most importantly for us,
\begin{itemize}
\item it is generically a smooth hyper-K\"ahler manifold;

\item it admits a Hamiltonian $S^1$-action, analogous to the familiar Hitchin action on the moduli space of unramified or tamely ramified Higgs bundles \cite{Hitchin:1986vp};

\item it has a projection onto an affine variety $\CB$, identified with the 4d Coulomb branch,
\be
\pi:\quad \CM_H \mapsto \CB;
\ee
\item the $S^1$ also acts on $\CB$, and $\pi$ is $S^1$-equivariant.
\end{itemize}

We will denote the three independent complex structures on $\CM_H$ as $I,J$ and $K$, and the three corresponding K\"ahler forms as $\omega_I,\omega_J,\omega_K$. The $S^1$-action is holomorphic in $I$, but acts non-trivially on $\Omega_I:=\omega_J+i\omega_K$ as
\be
\theta\in S^1:\quad \Omega_I\mapsto e^{N\cdot i\theta}\Omega_I,
\ee
where $N$ is expected to be the number of Stokes rays centered at the irregular singularity. Without wild ramification, $N$ equals one, and, as a consequence, the $S^1$-action is not a global symmetry of the theory but instead an R-symmetry. A special feature of the wild Hitchin moduli space is that now $N>1$, and a $\Z_N$ subgroup of $S^1$ preserves the hyper-K\"ahler structure of $\CM_H$ and becomes a discrete global symmetry. We will further assume that the action of $\Z_N$ on $\CB$ has no extra fixed points besides the origin $0\in\CB$. This is equivalent to the condition that no generators of the Coulomb branch spectrum have integral scaling dimensions, which is obeyed by all Argyres-Douglas theories studied in \cite{Fredrickson:2017yka,Fredrickson:2017jcf}. In fact, most results in this paper apply to more general 4d $\CN=2$ SCFTs that satisfy this ``non-integrality'' condition, even if their Coulomb branches cannot be realized as wild Hitchin moduli spaces.

As the $\Z_N$ global symmetry plays a crucial role in our construction, we now pause to make a few remarks about it.
\begin{itemize}

\item The $\Z_N$ fixed points are all isolated. This is because the nilpotent cone $\pi^{-1}(0)$ is Lagrangian, while connected components of the fixed locus are symplectic as $\Z_N$ preserves $\Omega$. By assumption, all components of the $\Z_N$ fixed locus belong to the nilpotent cone $\pi^{-1}(0)$, so they all have to be zero-dimensional. 

\item This further implies that the $S^1$ fixed points are exactly $\Z_N$ fixed points, as otherwise we would have a continuous $S^1$-orbit of $\Z_N$ fixed points. The wild Higgs bundles corresponding to fixed points in several infinite families of $\CM_H$ are explicitly constructed in \cite{Fredrickson:2017yka,Fredrickson:2017jcf}.

\item This $S^1$ isometry group of $\CM_H$ can be viewed as a covering of the $U(1)_{r}$ subgroup of the R-symmetry group of the Argyres-Douglas theory, and $\Z_N$ acts via deck transformations. This discrete global symmetry is used to turn on a holonomy in \eqref{M3M4} without breaking the supersymmetry. If the holonomy instead was not in the $\Z_N$ subgroup, it would not be possible to preserve supersymmetry along arbitrary three-manifolds, and would not lead to a TQFT.

\item After twisting a 4d $\CN=2$ theory on a Seifert manifold, the $U(1)_{\ft}$ becomes a flavor symmetry for the $\CN=2$ quantum mechanics in the remaining spacetime direction. As we expect states that contribute to the topological index to transform trivially under $SU(2)_R$, counting the $U(1)_r$ charge is the same as counting the $U(1)_{\ft}$ charge. We will follow the convention in \cite{Fredrickson:2017yka} for the fugacity $\frak{t}$ of $U(1)_{\ft}$. Namely, $\frak{t}$ takes values in the $N$-fold covering of the complex plane, branched at the origin. On this covering space, $\frak{t}=e^{2\pi i \gamma}$ is different from $\frak{t}=1$ as long as $\gamma \not\equiv 0 \pmod N$, and represents a non-trivial discrete holonomy. This is precisely the type of holonomies that we will turn on.

\item In the rest of the paper, when we encounter fractional powers of $\ft$ and make the substitution $\ft=e^{2\pi i \gamma}$, it will be understood that we stay on the first sheet, and $\ft^{\gamma/N}$ will be replaced with $e^{2\pi i \gamma/N}$. One might want to instead define $\ft':=\frak{t}^{1/N}$ to get rid of fractional powers and branch cuts, but the normalization we use has several benefits. For example, in this normalization, the powers of $\ft$ can be interpreted as scaling dimensions of operators, and having fractional scaling dimensions is a characteristic feature of Argyres-Douglas theories.

\item One consequence of the compactness of the $\Z_N$ fixed point set is that the substitution $\ft^{\gamma/N}=e^{2\pi i \gamma/N}$ does not lead to divergences if $\gamma \in \Z_N^\times$ is coprime to $N$. We restrict to such $\gamma$ henceforth.

\end{itemize}

\subsubsection*{The low energy effective theory}

Without the discrete holonomy, the low energy effective theory of a generalized Argyres-Douglas theory on $S^1 \times M_3$ is described by a sigma-model into $\CM_H$. Therefore, after the topological twist, one would naively expect to have a Rozansky-Witten theory with the target $\CM_H$~\cite{Rozansky:1996bq}. However, it has several undesirable properties:
\begin{itemize}
\item as $\CM_H$ is non-compact and not asymptotically flat, it is not clear whether the theory is well-define;
\item when $b_1(M_3)$ is large, one expects the partition function to vanish;
\item the action of the mapping class group MCG$(\Sigma)$ on the Hilbert space associated with $\Sigma$ factors through the group $\mathrm{Sp}(2g,\Z)$, making it less interesting.
\end{itemize}
We expect that all of these problems are gone with the discrete holonomy turned on:
\begin{itemize}
	\item non-compactness is not a problem as the set of $\Z_N$ fixed points is compact;
	\item the partition function no longer vanishes even when $b_1(M_3)$ is large, as can be seem in examples below;
	\item we now expect more interesting ``quantum representations'' of the mapping class group to appear, since many chiral algebras associated with generalized Argyres-Douglas theories can be constructed from quantum groups \cite{Xie:2016evu,Creutzig:2017qyf}.
\end{itemize}

Then, what is the true low-energy effective theory with the discrete holonomy turned on?
Using Morse theory, $\CM_H$ can be decomposed into copies of affine spaces glued together,
each identified with the normal bundle of an $S^1$ fixed point. Therefore, the low energy effective theory can be viewed as a collection of free theories coupled together. After topological twist and factoring out a tower of states, which will be explained more precisely later in the case of $M_3=L(p,1)$, each affine space will contribute one vacuum state to the TQFT. Therefore, one expects that the (isomorphism classes of) simple objects in the MTC are in 1-to-1 correspondence with the $S^1$ fixed points, and that structures of the MTC such as fusion rules, braiding, $S$ and $T$ matrices can be deduced from how the normal bundles to the fixed points are glued together to form $\CM_H$, or equivalently how the free theories are coupled together to form the true low-energy effective theory. The study of the low energy effective theory in the presence of discrete holonomy may be related to the discrete gauging of the $\Z_N$ symmetry briefly discussed in \cite{Argyres:2016yzz}. See also \cite{Buican:2015hsa} for a discussion in the case without holonomy.

\subsubsection*{The case of $M_3=L(p,1)$}

When $M_3$ is the lens space $L(p,1)$, supersymmetry can be preserved for generic values of $\frak{t}$, and the partition function is expected to compute the equivariant index of a line bundle on the Coulomb branch \cite{Gukov:2016lki}. For generalized Argyres-Douglas theories, such quantities are sometimes referred to as the ``wild Hitchin characters.'' For the $(A_1,A_2)$ theory, it is given by
\be
Z_{(A_1,A_2)}(\frak{t}) = \frac{1}{(1-\ft^{\frac{2}{5}})(1-\ft^{\frac{3}{5}})}+\frac{\ft^{\frac{p}{5}}}{(1-\ft^{\frac{6}{5}})(1-\ft^{-\frac{1}{5}})},
\ee
where the two terms come from the two $S^1$ fixed points in the wild Hitchin moduli space $\CM_H^{(A_1,A_2)}$. This theory has $\Z_5$ symmetry, and we can turn on a non-trivial holonomy by setting $\ft=e^{2\pi i}$. After this substitution, as one can easily verify, $Z_{(A_1,A_2)}$ indeed agrees with $(ST^pS)_{0,0}$ with $S$ and $T$ of the Lee-Yang model given by \eqref{ST}, up to a phase factor. In fact, such agreement was checked for three infinite families of generalized Argyres-Douglas theories in \cite{Fredrickson:2017yka}. For example, we have the following equality for $(A_1,A_{2n})$ theories,
\be
\left(S T^p S\right)_{0,0} = e^{2\pi i p \mu_0} Z_{(A_1, A_{2n})}(\ft=e^{2\pi i}),
\ee
if one uses the $S$ and $T$ matrices of the $(2,2n+3)$ minimal model, which was conjectured to be the chiral algebra associated with the $(A_1,A_{2n})$ theory \cite{Cordova:2015nma}. Here $\mu_0=\frac{1}{24}-\frac{1}{8(2n+3)}$ is independent of $p$ and can be eliminated by shifting the $S^1$ moment map.

There is evidence that this relation also holds for $(A_{m-1},A_{n-1})$ theories with $n$ co-prime to $m>2$. For example, the values of the $S^1$ moment map at fixed points are related to the eigenvalues of the $T$ matrix in the corresponding chiral algebras \cite{Fredrickson:2017jcf}.

In the next section, we check our proposed relation between MTC and partition functions of Argyres-Douglas theories by testing it on a much larger class of 3-manifolds. Before that, we emphasize again the importance of the discrete holonomy $\gamma\in \Z_N^\times$ by explaining the role it played in the case of $M_3=L(p,1)$.  

For more general Argyres-Douglas theories, it is still true that the normal bundle to a fixed point $\lambda$ will contribute to the Floer-like homology ({\it i.e.}~the $Q$-cohomology of the twisted 4d theory on $L(p,1)$) a vacuum with a tower of states $\CT^+_{\lambda}$ attached to it. The $U(1)_{\ft}$ character of every such tower has the form
\be
\chi_{\ft} (\CT^+_{\lambda}) \; = \; \frac{\ft^{n_{\lambda}}}{E_{\lambda} (\ft)}
\label{Ttdim}
\ee
where $n_\lambda$ is the charge of the vacuum and $E_{\lambda} (\ft)$ is a polynomial of degree $\dim_{\mathbb{C}} \CM_H$.
What makes Argyres-Douglas theories special is that \eqref{Ttdim}
has a finite value at $\ft=e^{2\pi i \gamma}$, which can be regarded as a regularization
of the power series in $\ft$ that encodes the graded dimensions of $\CT^+_{\lambda}$.
Therefore, while states $\vert \lambda \rangle$ in our 3d TQFT on $\mathbb{R} \times T^2$
are in 1-to-1 correspondence with $S^1$ fixed points on $\CM_H$,
the normal bundles to the fixed points and the corresponding towers $\CT^+_{\lambda}$ do play an important role;
namely, their regularized $U(1)_{\ft}$ character encodes information about the $S$ and $T$ matrices.

\subsection*{RG flows from 4d $\CN=1$ theories}

While most Argyres-Douglas theories are believed to be non-Lagrangian,
one might reach them by RG flows starting from 4d $\CN=1$ Lagrangian theories.
In many examples, candidates for such UV $\CN=1$ Lagrangians (modulo decoupling of free fields that can be easily accounted for) were conjectured in \cite{Maruyoshi:2016tqk,Maruyoshi:2016aim,Agarwal:2016pjo,Benvenuti:2017lle,Benvenuti:2017bpg,Evtikhiev:2017heo}.

As topologically twisted partition functions $Z (S^1 \times M_3)$ are RG invariant, the $\CN=1$ Lagrangian description suffices to compute them for a particular class of 3-manifolds, such as $M_3 = S^1 \times \Sigma$ or more general Seifert manifolds. In a related context, such computations were done in \cite{Gukov:2015sna,Benini:2015noa,Benini:2016hjo,Gukov:2017zao,Closset:2017bse,Closset:2018ghr}, which we closely follow here.

To keep the presentation simple, we will consider three Argyres-Douglas theories: the $(A_1,A_2)$, $(A_1,A_3)$, and $(A_1,D_3)$ theories. The latter two theories are conjectured to be identical, and indeed we find that their partition functions always agree and give rise to the same MTC.

Denote the total space of a degree-$p$ circle bundle over a genus-$g$ Riemann surface as $L_g(p)$. We use localization to compute the $S^1 \times L_g(p)$ twisted partition functions of 4d $\CN=1$ theories that
flow to the desired Argyres-Douglas theories. In this case, the general formula \eqref{ZSTgeneral} reduces to \footnote{In general, the 3d TQFT determined by an MTC has anomaly, and the partition function will depend on a choice of framing. Here, we have used the ``Seifert framing,'' in which the partition function on $L_g(p)$ takes the cleanest form.}
\be
Z_{\text{AD}} (T^2 \times \Sigma_g) \; = \; \sum_{\lambda} (S_{0 \lambda})^{2-2g} T_{\lambda\lambda}^p,
\ee
where we have used the fact that the $T$ matrix is diagonal in the basis given by simple objects of the MTC. Once the left-hand side is computed by standard localization techniques in the UV theory,
we can easily extract the $S$ and $T$ matrices using \eqref{Shandle} and \eqref{Tfibering} combined with modularity.

Lagrangians of 4d $\CN=1$ theories that we use all have a $U(1)_\CF$ flavor symmetry, which at the IR fixed point is embedded into the $\CN=2$ R-symmetry as $\CF=R-r$, with $r$ a generator of $U(1)_r$, and $R$ a Cartan generator of $SU(2)_R$. It plays a crucial role in our computations: turning on holonomy for this $U(1)_\CF$ implements the desired holonomy for $U(1)_{\ft}$ in the IR. This works for $M_3=L_g(p)$ and, possibly, for other Seifert manifolds. Different methods are required to compute partition functions on more general $M_3$. One practical way (which relies on the TQFT existence)  is to use the MTC, instead of applying localization to the 4d $\CN=1$ Lagrangian theory.

\subsubsection*{$(A_1, A_2)$ theory}
We use the $\CN=1$ Lagrangian of \cite{Maruyoshi:2016aim,Maruyoshi:2016tqk,Agarwal:2016pjo,Benvenuti:2017lle,Benvenuti:2017bpg,Evtikhiev:2017heo} that flows to the $(A_1, A_2)$ theory. The UV description is an $SU(2)$ gauge theory with the matter content summarized in Table \ref{tab:A1A2}, along with a superpotential
\be
W=q \phi q+u q'\phi q'.
\ee
This $\CN=1$ theory has a flavor symmetry $U(1)_{\CF}$ that will become the $U(1)_{\ft}$ subgroup of the $\CN=2$ R-symmetry in the infrared. So we use $\ft$ for the holonomy of this flavor symmetry.

\begin{table}[htb]
	\be
	\begin{array}{l@{\;}|@{\;}cccc}
		& q & q' & \phi & u \\\hline
		SU(2)_{\text{gauge}} & \Box & \Box & \text{adj} & 1 \\
		U(1)_R & 1 & 1 & 0 & 0 \\
		U(1)_\CF & \frac15     & \frac75      & -\frac25    & -\frac{12}5
	\end{array}		\notag \ee \caption{Field content of the 4d $\CN=1$ Lagrangian theory that flows to the $(A_1,A_2)$ Argyres-Douglas theory.}\label{tab:A1A2}
\end{table}

In the localization computation, one has an additional parameter $\tau$ coming from the complex structure of a $T^2$ whose longitude and meridian are, respectively, the $S^1$ in \eqref{M3M4} and the Seifert fiber of $M_3$.  The topologically twisted partition function does not depend on $\tau$ and is conveniently recovered in the $\tau\rightarrow 0$ limit (which is also equivalent to the $\tau \rightarrow i\infty$ limit by modular invariance). After a straightforward but technical computation that involves solving the Bethe ansatz equation in this limit (and taking care of the ``holonomy saddles'' \cite{Gross:1980br,Hwang:2018riu}), we find the following result.

For $M_3=L_g(p)$, the partition function is given by
\be
Z_{(A_1,A_2)}(S^1\times {L_g(p)};\ft)=\sum_i (\CH_i)^{g-1} (\CF_i)^p.
\ee
Here the sum goes over the solutions to Bethe equations, while $\CH_i$ and $\CF_i$ are the VEVs of the ``handle-gluing'' operator and the ``fibering'' operator at the $i$-th Bethe vacuum respectively. They all are functions of $\ft$.

For the $\CN=1$ theory flowing to $(A_1,A_2)$, there are two Bethe vacua, and we have
\begin{align}
\CH_1&=\ft^{1/10} + \ft^{-1/10} - \ft^{1/2} - \ft^{-1/2},\cr
\CH_2&=\ft^{7/10} + \ft^{-7/10} -\ft^{1/2} - \ft^{-1/2},\cr
\CF_1&=\ft^{-1/60},\qquad \CF_2=\ft^{11/60}.
\end{align}
They can be assembled into ``$\ft$-dependent $S$ and $T$ matrices,'' 
\begin{align}
T(\ft)=\left(\begin{matrix}
\CF_2 & 0\cr
0 & \CF_1
\end{matrix} \right),\quad S(\ft)=\left(\begin{matrix}
-\CH_2^{-1/2} & \CH_1^{-1/2}\cr
\CH_1^{-1/2} & \CH_2^{-1/2}
\end{matrix} \right).
\end{align}
Let us enumerate rows and columns by $\lambda=0,1$. Then the partition function on $L_g(p)$ can be written as
\begin{align}
Z_{(A_1,A_2)}(\ft)=(\CH_1)^{g-1} (\CF_1)^p+(\CH_2)^{g-1} (\CF_2)^p\nonumber\\
=\sum_{\lambda=0,1}(S_{0\lambda}(\ft))^{2-2g}(T_{\lambda\lambda}(\ft))^p.
\end{align}

We would like these $S$ and $T$ to be modular matrices. Namely, they should obey $S^2 = (ST)^3=C$, where $C$ is the charge conjugation matrix in the MTC satisfying $C^2=1$. A computation shows that
\begin{align}
S(\ft)^2 &= -\frac{\ft^{1/2}}{\ft^{1/5}+\ft^{2/5}+\ft^{3/5}+\ft^{4/5}+\ft+1} \times {\mathbf 1}_{2\times 2},\cr
\left(S(\ft)T(\ft)\right)^3 &= \frac{\left(\ft^{1/5}-1\right) \ft^{13/20}}{\sqrt{\ft^{7/10}+{\ft^{-7/10}}-\ft^{1/2}-\ft^{-1/2}}} \cr
& \times \frac{1}{\ft^{1/5}+\ft^{2/5}+\ft^{3/5}+\ft^{4/5}+\ft+1} \times {\bf 1}_{2\times 2}.
\end{align}
It is easy to verify that they are modular precisely at $\ft=e^{2\pi i \gamma}$ with $\gamma\in \Z_5^\times$. For $\gamma=1$, they indeed agree with the expressions in \eqref{ST}. 

A one-parameter family of $S$ and $T$ matrices is also constructed in \cite{Kozcaz:2018usv}, which is related to our $S(\ft)$ and $T(\ft)$ by overall factors. Although the entire family is modular, due to Ocneanu rigidity, one doesn't expect an MTC to exist for generic members of this family.

\subsubsection*{$(A_1, A_3)$ and $(A_1, D_3)$ theories}

We will now check our proposal for the $(A_1, A_3)$ and $(A_1, D_3)$ theories. They have an additional $SO(3)$ flavor symmetry, for which we can turn on a fugacity $\frak{s}$. The field contents of the $\CN=1$ UV descriptions are given in Tables~\ref{tab:A1A3} and~\ref{tab:A1D3}. Although the two theories are conjectured to be identical in the IR, two UV descriptions given here are different. For example, only a $U(1)_B\subset SO(3)$ flavor symmetry is visible in the UV Lagrangian of the $(A_1,A_3)$ theory.

Each of the two theories has three Bethe vacua, and hence three VEVs of the handle-gluing operator $\CH_i(\ft,\fs)$, $i=1,2,3$ and three VEVs of the fibering operator $\CF_i(\ft,\fs)$, $i=1,2,3$, all of which are functions of $\ft$ and $\fs$. They determine some entries of the modular matrices $S$ and $T$, while other entries should be fixed using $S^2=(ST)^3=C$, with $C^2=1$.

\begin{table}[htb]
\be
	\begin{array}{l@{\;}|@{\;}cccc}
		& q & \tilde q & \phi & u \\\hline
		SU(2)_{\text{gauge}} & \Box & \Box & \text{adj} & 1 \\
		U(1)_B & 1 & -1 & 0 & 0 \\
		U(1)_{R} & 1 & 1 & 0 & 0 \\
		U(1)_\CF &   \frac43     &   \frac43     & -\frac23   & -\frac83
	\end{array}
	\notag \ee
	\caption{Field content of 4d $\CN=1$ Lagrangian theory that flow to the $(A_1,A_3)$
		Argyres-Douglas theory. The superpotential is $W=uq \tilde{q}.$}
	\label{tab:A1A3}
\end{table}

\begin{table}[htb]
	\be
	\begin{array}{l@{\;}|@{\;}cccc}
		& q & \tilde q & \phi & u \\\hline
		SU(2)_{\text{gauge}} & \Box & \Box & \text{adj} & 1 \\
		SO(3)_{\text{flavor}}&  3   &  1   &  1         & 1 \\
		U(1)_{R} & 1 & 1 & 0 & 0 \\
		U(1)_\CF &   \frac13     &   \frac53     & -\frac23   & -\frac83
	\end{array}
	\notag \ee
	\caption{Field content of a 4d $\CN=1$ Lagrangian theory that flows to the $(A_1,D_3)$
		Argyres-Douglas theory. The superpotential is $W=q\phi q + u\tilde{q}\phi\tilde{q}$.} \label{tab:A1D3}
\end{table}

For generic values of $\fs$ \footnote{For general $M_3$, we expect the excitations along the directions of the Higgs branch to be frozen after the topological twist, as such modes become spinors on $M_3$ but there are no solutions for harmonic spinors on $M_3$ generically. However, as our $M_3$ is a Seifert manifold with a very non-generic metric, there can be such solutions. We believe the role played by choosing a generic value of $\fs$ is to eliminate such solutions.} (which in practice can be chosen to be any values besides $\pm1$ and $\pm i$), the modularity holds again at $\ft=e^{2\pi i}$. We find for both theories~\footnote{In this case, we actually have 4 solutions that are related by the $\Z_3^\times$ action and tensoring with an ``almost trivial'' rank-1 MTC with $S=-1$. The latter operation gives a pair of MTCs that contain essentially the same information, and we will not distinguish between them. We have also identified MTCs obtained by tensoring with the invertible 3d TQFT given by the $(E_8)_1$ Chern-Simons theory.
}
\begin{align}
\CH_1&=\CH_2=\CH_3=3,\cr
\CF_1&=e^{i\pi/2},\cr
\CF_2&=\CF_3=e^{-i\pi/6}.
\end{align}
Such values give answers that perfectly agree with modular matrices acting on three admissible representations of $\mathfrak{su}(2)_{-4/3}$, which is conjectured to be the chiral algebra corresponding to the $(A_1, D_3)$ theory. Namely,
\be
Z_{(A_1,D_3)}(T^2\times \Sigma_g) = 3^g=\sum_{\lambda=0}^2 (S_{0\lambda})^{2-2g},
\ee
and
\begin{multline}
Z_{(A_1,D_3)}(S^1\times L_g(p)) 
= 3^{g-1} (e^{p i\pi/2} +  e^{-p i\pi/6} + e^{-p i\pi/6})  \\
=\sum_{\lambda=0}^2 (S_{0\lambda})^{2-2g}(T_{\lambda\lambda})^p,
\end{multline}
where 
\begin{align}
S&=-\frac1{\sqrt{3}}\left(\begin{matrix}
1  & -1 & 1\cr
-1 & \epsilon^2 & -\epsilon\cr
1  & -\epsilon  & \epsilon^2
\end{matrix} \right) \text{  with } \epsilon=e^{2\pi i/3},\cr
T&=\left(\begin{matrix}
e^{i\pi/2}  & 0 & 0\cr
0 & e^{-i\pi/6} & 0\cr
0  & 0 & e^{-i\pi/6}
\end{matrix} \right).
\end{align}
One interesting feature of this MTC is that it is not related to a unitary MTC via action of the Galois group $\Z_3^\times$. Instead, $\CC_{(A_1,D_3)}^1$ and $\CC_{(A_1,D_3)}^2$ are non-unitary MTCs that are complex conjugate to each other. However, if one ``flips the sign'' for the object labeled by $\lambda=1$, one can get the $\Z_3$ MTC, which can be realized by either the $SU(3)_1$ or $(E_6)_1$ Chern-Simons theory \cite{RSW}.    

\subsection*{Future directions}

It would be interesting to explore twisted partition functions on $S^1\times M_3$ for more general $M_3$ and more general 4d $\CN=2$ theories, to understand more fully the structure of the MTCs proposed here, the Galois action, and their interplay with physics of SCFTs. 

The bijection between simple objects in the MTC and $S^1$-fixed points in $\mathcal M_H$ \cite[Sec.~5]{Fredrickson:2017yka} reminds us the bijection between simple modules in the category $\mathcal O$ of a quantized Coulomb branch and $S^1$ fixed points in the resolved Higgs branch for a 3d $\mathcal N=4$ gauge theory via Hikita conjecture. (See \cite{KTWWY} for the special case when the gauge theory is associated with a quiver of type ADE.)  A generalized AD theory discussed here has a quiver gauge theory as its 3d mirror in many cases \cite{Nanopoulos:2010bv}, and its quantized Coulomb branch is close to Zhu's algebra of the chiral algebra. (See \cite{Beem:2017ooy} for a related discussion about chiral algebra and the 4d Higgs branch.) Part of simple objects in the MTC should correspond to simple modules of the Zhu's algebra, and hence have something to do with $S^1$ fixed points in the resolved Higgs branch, which is a quiver variety. This restriction of the bijection should explain an observation in \cite[Sec.~4.2]{Fredrickson:2017yka} that part of the $S^1$ fixed points are coming from quiver varieties. We also wish to better understand the role of mass parameters in the bijection in \cite{Fredrickson:2017yka}, which, identified as FI parameters in the 3d mirror theory, are used to define the category $\mathcal O$ of quantized Coulomb branch and the resolution of Higgs branch in \cite{KTWWY}.
\\

\begin{acknowledgements}
	We thank J.E.Andersen, B.Feigin, L.Fredrickson, K.Maruyoshi and N.Nekrasov for interesting discussions. The work of MD, SG, DP and KY was supported by the Walter Burke Institute for Theoretical Physics and the U.S. Department of Energy, Office of Science, Office of High Energy Physics, under Award No.\ DE{-}SC0011632. The work of MD was also supported by the Sherman Fairchild Foundation. The work of SG was also supported by the National Science Foundation under Grant No.~NSF DMS 1664240. The work of DP was also supported in part by the center of excellence grant ``Center for Quantum Geometry of Moduli Space" from the Danish National Research Foundation (DNRF95). The research of HN was supported in part by the World Premier International
	Research Center Initiative (WPI Initiative), MEXT, Japan, and by JSPS
	Grant Number 16H06335.
\end{acknowledgements}


\end{document}